\documentclass[journal]{IEEEtran}
\usepackage{xcolor}
\usepackage{graphicx}
\usepackage{pgfplots}
\pgfplotsset{compat=newest}
\usepackage{cite}
\usepackage{subfigure}
\usepackage{lipsum}
\usepackage{stfloats}

\usepackage{comment}
\usepackage{xcolor}
\specialcomment{W2Cut}{ \color{orange!30!yellow}}{ \color{black}}
\excludecomment{W2Cut}
\begin{document}
\newcommand{\Mori}[1]{\textcolor{red}{ {#1} }}
\newcommand{\DAtt}[1]{\textcolor{red}{ {#1} }}

\newcommand{\FigSize}{0.15}
\newcommand{\FigSizeFF}{0.14}

\graphicspath{{Figs_PDF/}}

\title{Point contacts in modeling conducting 2D planar structures}

\author{David~V.~Thiel~\IEEEmembership{Senior~Member,~IEEE,}, Morteza~Shahpari~\IEEEmembership{Student~Member,~IEEE},      Jan~Hettenhausen,
Andrew~Lewis~\IEEEmembership{Member,~IEEE,} 
\thanks{This work is partly funded by a grant from Australian Research Council DP130102098.}
\thanks{D. V. Thiel and M. Shahpari are with the School of Engineering, Griffith University, Nathan, Queensland 4111, Australia}
\thanks{J. Hettenhausen and A. Lewis are with the Institute for Integrated and Intelligent Systems, Griffith University, Queensland, Australia }
}

\markboth{IEEE Antennas and Wireless Propagation Letters, Vol.14, 2015} {righthead}
\maketitle

\begin{abstract}
Use of an optimization algorithm to improve performance of antennas and electromagnetic structures usually ends up in planar unusual shapes.
Using rectangular conducting elements the proposed structures sometimes have connections with only one single point in common between two neighboring areas.
The single point connections (\emph{point crossing}) can affect the electromagnetic performance of the structure.
In this letter, we illustrate the influence of point crossing on dipole and loop antennas using {MoM, FDTD, and FEM solvers}.
Current distribution, radiation pattern, and impedance properties for different junctions are different.
These solvers do not agree in the modeling of the point crossing junctions which is a warning about uncertainty in using such junctions.
However, solvers agree that a negligible change in the junction would significantly change the antenna performance.

We propose that one should consider both bridging and chamfering of the conflicting cells to find optimized structures.
This reduces the simulation time by 40\% using FDTD modeling, however no significant reduction is obtained using the MoM and FEM methods.


\end{abstract}

\section{Introduction}

\IEEEPARstart{O}{ptimization} algorithms help engineers to find the best possible set of designs.
Relying on the validity of the computing algorithms has some drawbacks as not all the proposed optimized structures are feasible to manufacture.
For instance, optimized 2D antenna structures with a ground plane were reported in \cite{Cismasu_2014_TAP,Cismasu_2014_AWPL}.
Their structure consisted of small PEC squares and the optimization routine eliminated squares randomly from the plane optimized for minimum $Q$.
Each new design was solved using method of moments code (MoM) \cite{Peterson_b1998}. 
The $Q$-factor was calculated with single frequency simulations.
Some of the proposed structures  have junctions with a single common point diagonally connecting adjacent PEC squares.

This is not the first time that optimization algorithms suggest structures which have neighbor squares with only one common point. 
For example a similar point crossing is reported for frequency selective surfaces \cite{Kern_2003_MWOPL,Kern_2005_TAP,Xia_2013_PIER,Bayraktar_2013_TAP}.
In some other works the neighbor areas overlap and make a path for the flow of current \cite{John_2007_AWPL,Kerkhoff_2007_TAP}.

Shahpari et. al \cite{Shahpari_2014_TAP} reported optimized meander line antennas based on the ant colony algorithm when radiation efficiency \cite{Galehdar_2009_IJRFCAD} was involved in the Pareto optimized front. 
These models included finite conductivity and finite width conductors.
For meander line antennas in \cite{Shahpari_2014_TAP,Galehdar_2009_IJRFCAD} this uncertainty did not exist because continuous wires were used which  could not fold and join themselves.

In this letter, we illustrate the effect of point crossing on the radiating systems.
It is shown that such junctions between perfect electric conducting squares behave like an ideal open circuit in MoM simulations, but imitate a short circuit in FEM and FDTD modeling.
However, small displacement of the cells would significantly change the flow of current over the junction and consequently change the impedance and radiation pattern of the antenna.
We propose that one should avoid point crossing, but should model chamfer and bridge connections to make sure that available optimization space are properly explored.
Simulation time of the different junctions with different EM solvers are compared in the section.~\ref{Sec_SimTime}.

\section{Effect of Point Crossing on antennas}

\begin{figure}
	\centering
	\subfigure[]{
		\includegraphics[width=0.6\linewidth, ]{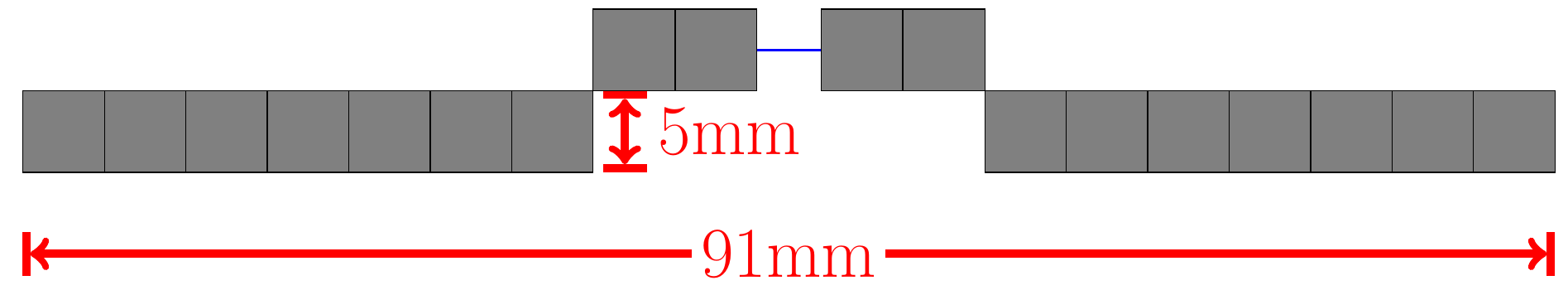}
		\label{Fig_Dipole}
	}\hfil
	\subfigure[]{
		\includegraphics[width=0.3\linewidth]{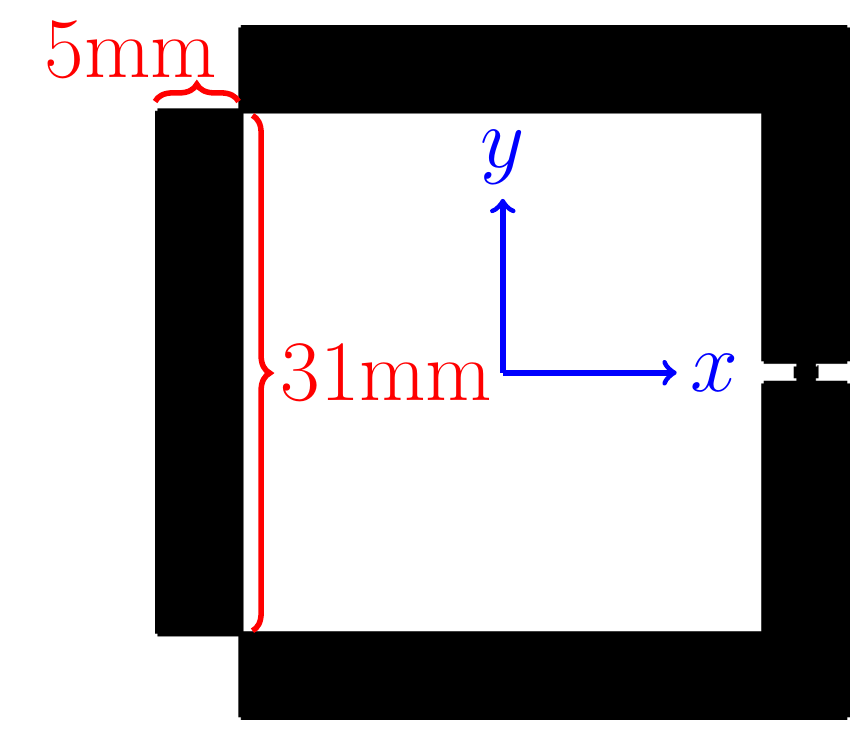}
		\label{Fig_SQLoop}
	}
	\caption{(a) Dipole antenna with a bend in each arm (b) Square loop antenna with two point crossing issues}
	\label{Fig_Dip_SQLoop}
\end{figure}
\begin{figure}
\includegraphics[width=\linewidth]{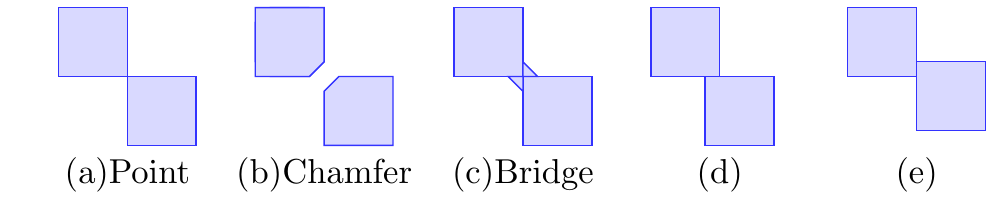}
\caption{Different scenarios for neighbor cells joining together}
\label{Fig_CornerShapes}
\end{figure}

\begin{figure}[t]
\includegraphics[width=\linewidth]{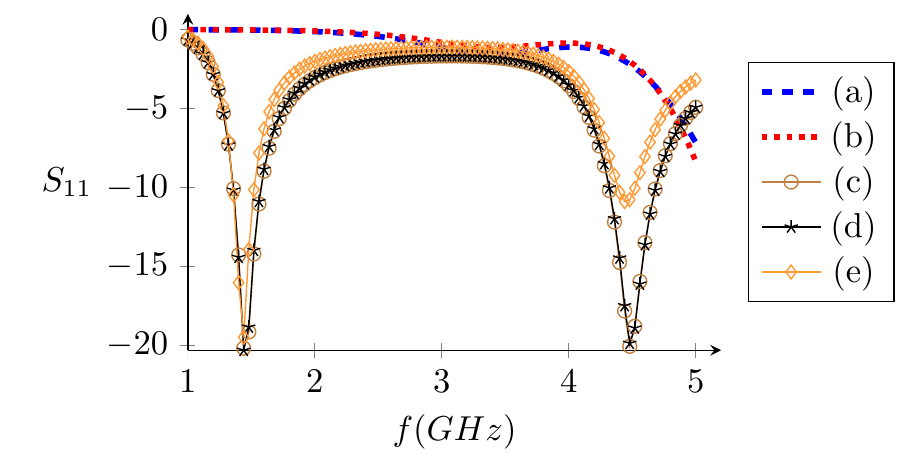}
\caption{Scattering parameter $(S_{11})$ of a bent dipole with junctions illustrated in Fig.\ref{Fig_CornerShapes} extracted from MoM simulations}
\label{Fig_Dipole_S11_Junction}
\end{figure}

\begin{figure*}[b]
\includegraphics[width=\linewidth]{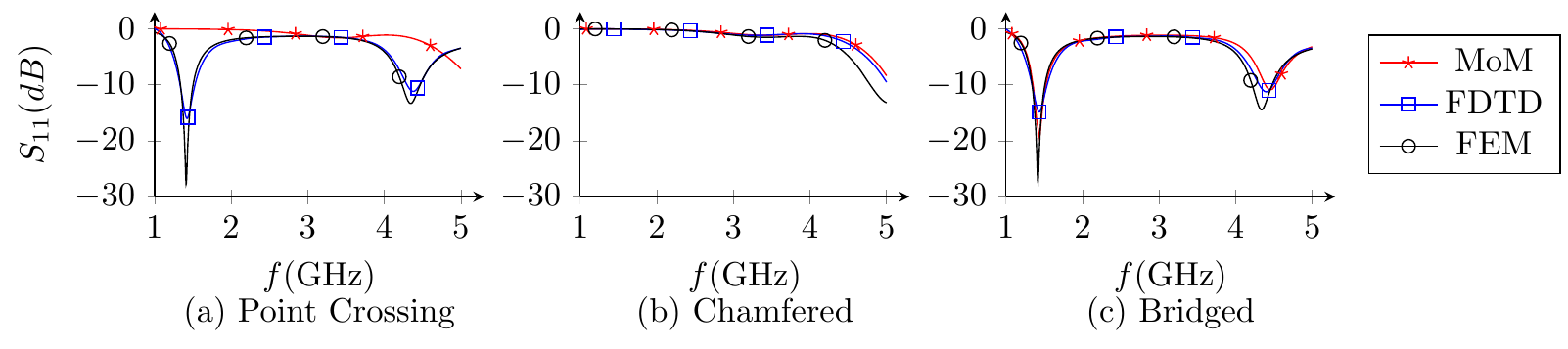}
\caption{Comparison of the $S_{11}$ of dipole with point crossing, chamfer and bridge junctions simulated with MoM, FDTD, and FEM methods.}
\label{Fig_Dipole_S11_Solver}
\end{figure*}

\begin{figure*}[b]
\centering
\includegraphics[width=0.95\linewidth]{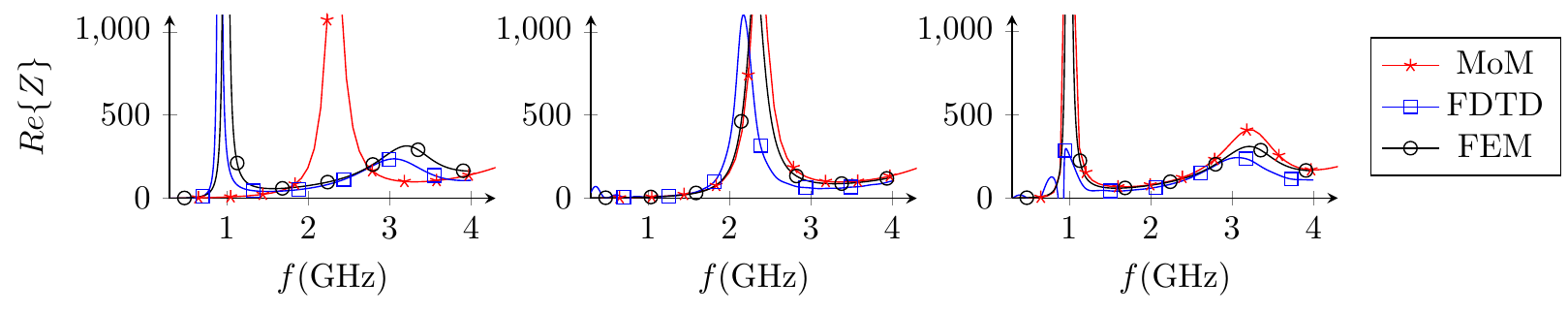}\\
\includegraphics[width=0.95\linewidth]{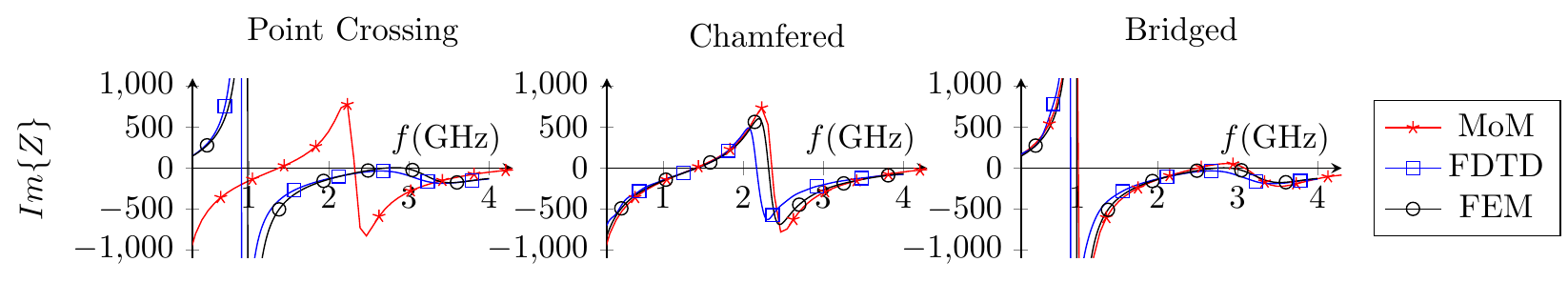}
\caption{Real and imaginary parts of the impedance of the loop antenna with point, chamfer, and bridge junctions simulated with MoM, FDTD, and FEM methods.}
\label{Fig_SQLoop_Imp}
\end{figure*}

\subsection{Example 1: dipole antenna}

We selected a simple dipole-like structure as the first example to investigate the issue of point crossing.
The structure is constructed by aligning rectangular cells in two rows which have only one point in common (see Fig.~\ref{Fig_Dipole}).
To examine the effect of the point crossing, we modeled other configurations including chamfering the conflicting cell, negligible shift in length and width of either of cells, or throughly connecting neighbor cells.
A single point contact between two cells cannot be fabricated as this is a connection point with an infinitely small cross-section between two perfectly conducting planar elements.
The alternate configurations in Fig.~\ref{Fig_CornerShapes} can be considered as the potential solutions of a real junction which due to manufacturing inaccuracies cells are either shifted, or trimmed, etc.
These structures were solved using a commercial MoM package \cite{FEKO}. 
Fig.~\ref{Fig_Dipole_S11_Junction} compares $S_{11}$ for the five slightly different dipole structures.
Assuming the ideal point crossing case (a) as a reference, there is little difference with elements that are completely separated but major differences occur when the squares overlap.



The effect of two PEC squares connected by a single point presents uncertainty about current flow. 
We modeled the dipole in Fig.~\ref{Fig_Dip_SQLoop} with MoM, FDTD, and FEM solvers. 
As illustrated in Fig.~\ref{Fig_Dipole_S11_Solver}, FDTD and FEM predict different results for the point crossing junction compared to the MoM method.
All 3 solvers give similar results for chamfered and bridged cases which are more practical in the real experiments.

\subsection{Example 2: loop antenna}

\begin{figure*}[t]
\centering
\includegraphics[width=1cm]{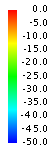}
\subfigure[Junction (a)]{
\includegraphics[width=\FigSize\textwidth]{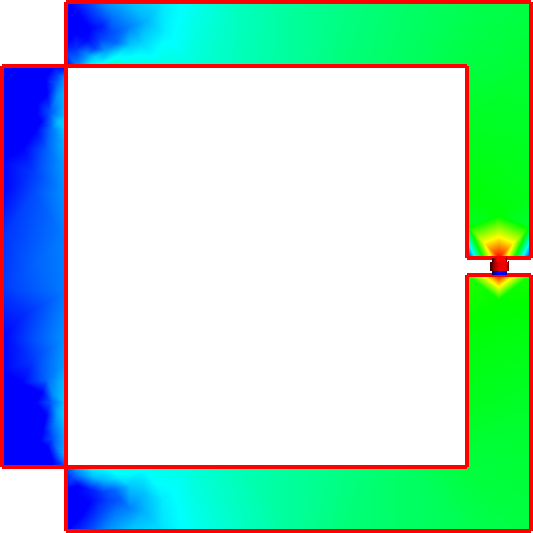}
}
\subfigure[Junction (b)]{
\includegraphics[width=\FigSize\textwidth]{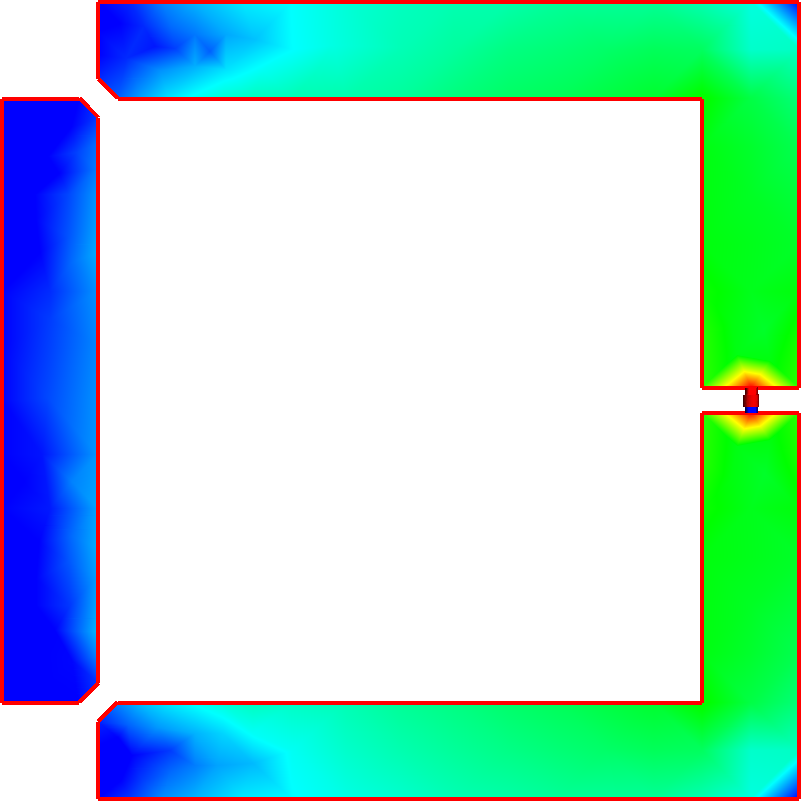}
}
\subfigure[Junction (c)]{
\includegraphics[width=\FigSize\textwidth]{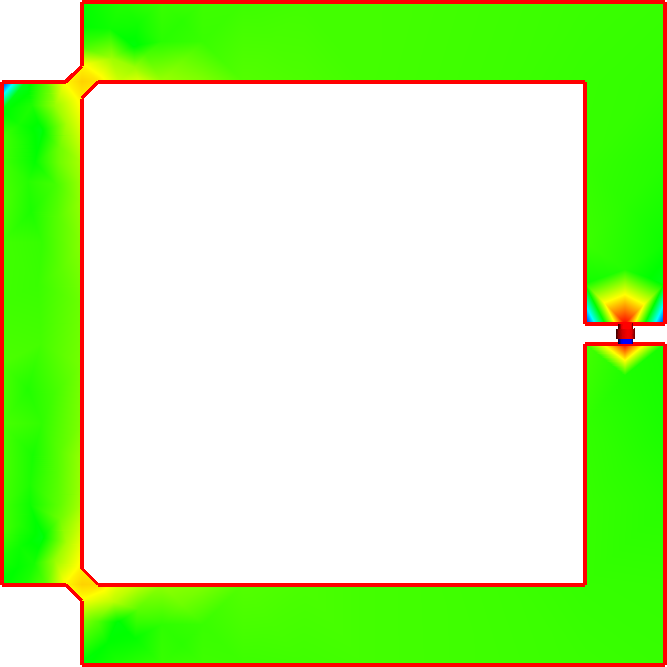}
}
\subfigure[Junction (d)]{
\includegraphics[width=\FigSize\textwidth]{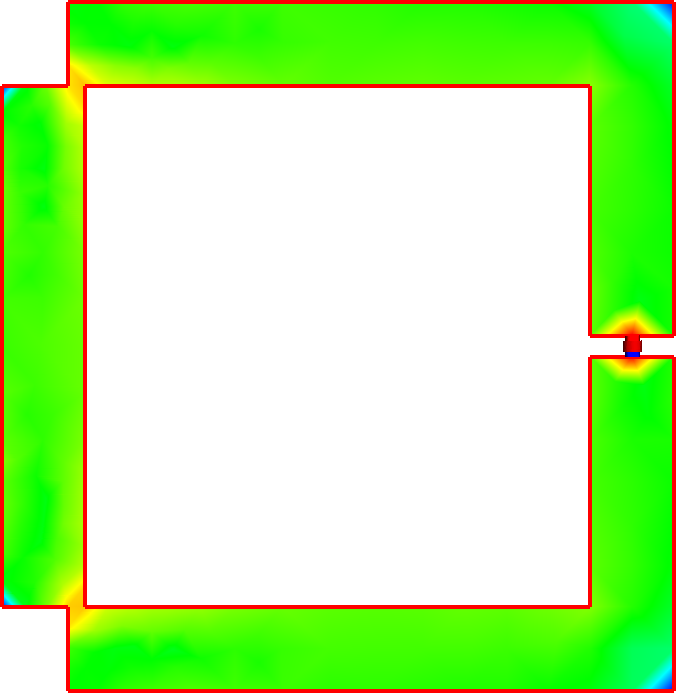}
}
\subfigure[Junction (e)]{
\includegraphics[width=\FigSize\textwidth]{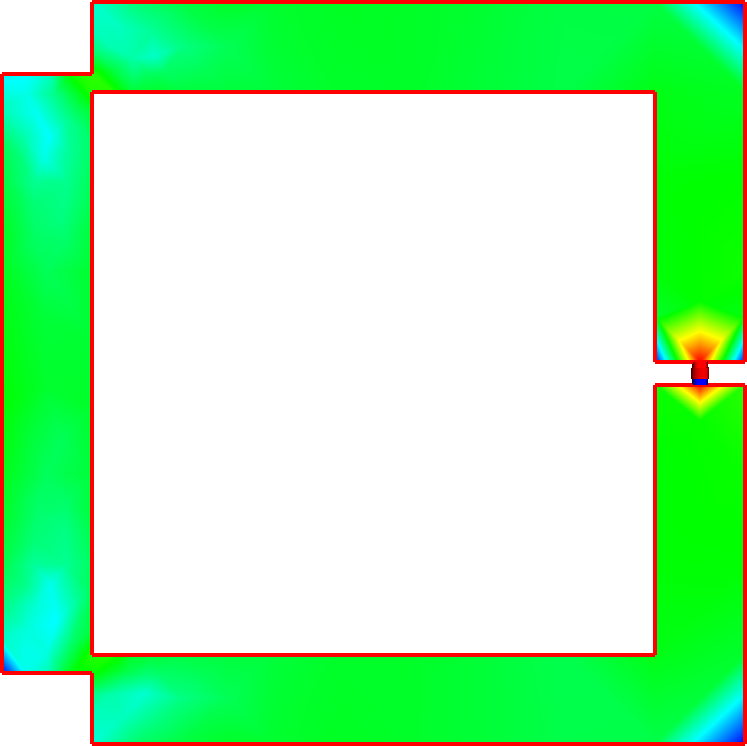}
}
\caption{Normalized magnitude (in $dB$) of current distribution of a square loop with junction types illustrated in Fig.\ref{Fig_CornerShapes} at $500$MHz (MoM simulation results).
}
\label{Fig_SQLoop_J}
\end{figure*}

A loop antenna was investigated to show the impact of point crossing junctions on the impedance $Z$, current distribution $J$, and the far field radiation.
We considered a square loop antenna which has point crossing at two neighbor vertices and is split into one U part and one straight line (see Fig~.\ref{Fig_SQLoop}).

Fig.~\ref{Fig_SQLoop_Imp} compares the real and imaginary parts of the impedance of the square loop with point crossing, chamfered, and bridge junctions, which are simulated with MoM, FDTD, and FEM methods.
The loop with chamfered junctions resonates around $1.3$GHz, and then changes to antiresonance around $2.2$GHz.
The bridged junction has inductive impedance properties (like a small loop) which shows anti-resonant and resonant behaviors at $1$GHz and $2.45$GHz, respectively.
The FDTD and FEM methods are not in agreement with MoM when modeling the loop with point crossing junction.
It is seen that MoM modeling does not allow the current flow through the junction, while FDTD and FEM allow that.
We believe that this disagreement is not a problem in the MoM, as one never can measure such point crossing junctions.
However, the significant shift from a chamfer to bridge is of particular importance.

\begin{figure}
\centering
\subfigure[ ]{
\includegraphics[height=\FigSizeFF\textwidth]{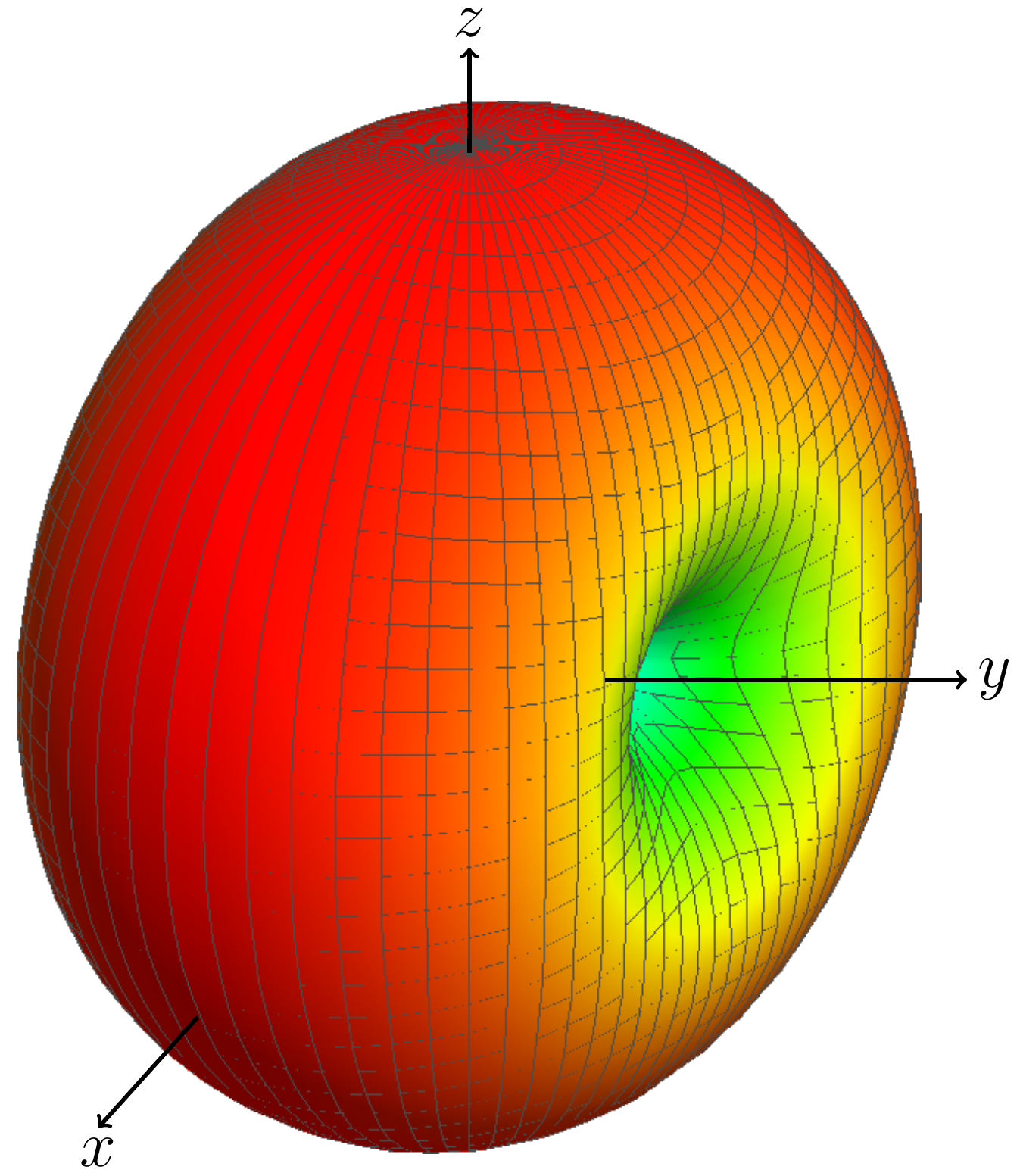}
}
\subfigure[ ]{
\includegraphics[width=\FigSize\textwidth]{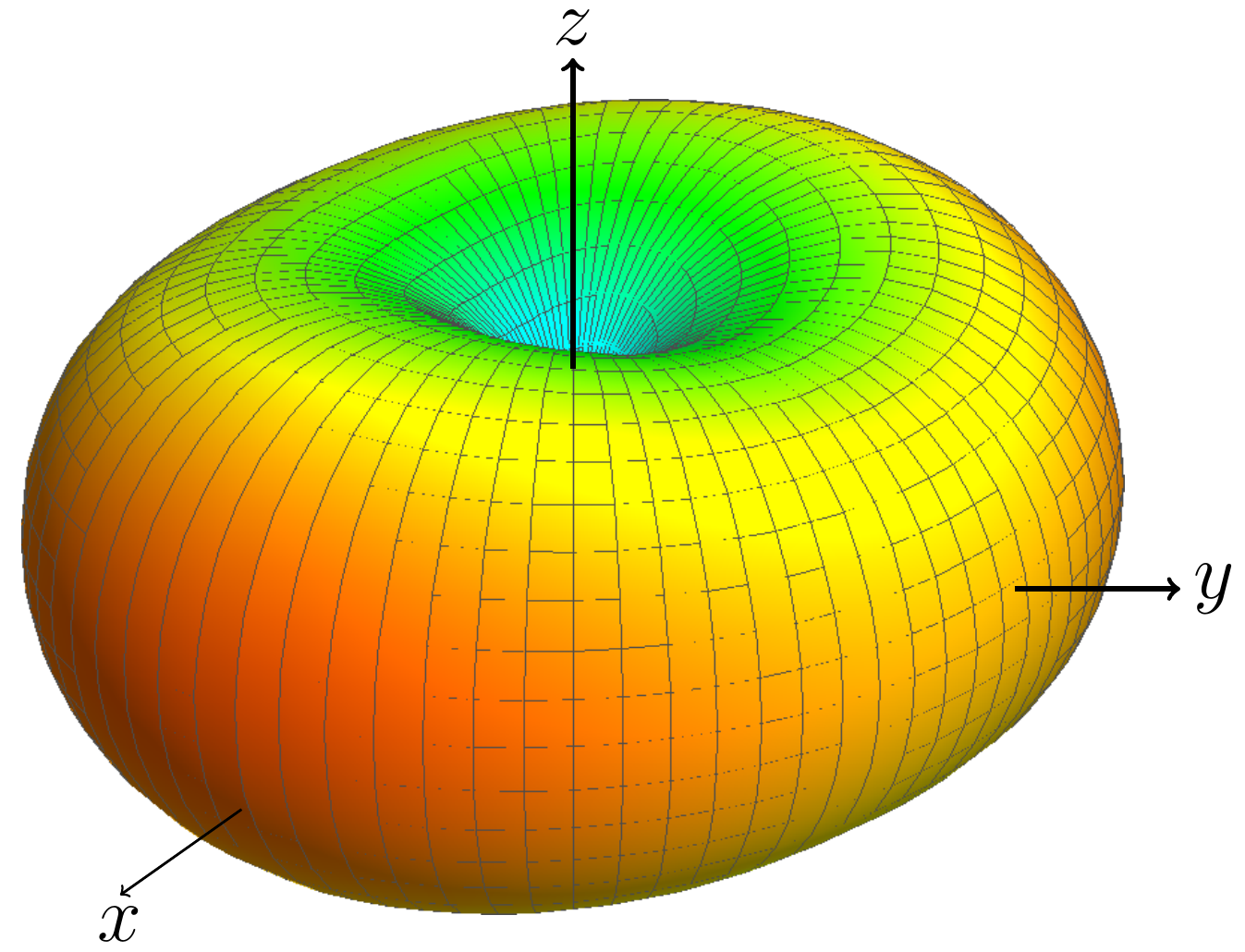}
}
\caption{Radiation pattern of a square loop with junction types illustrated in Fig.\ref{Fig_CornerShapes} at $500$MHz}
\label{Fig_SQLoop_FF}
\end{figure}

Fig,~\ref{Fig_SQLoop_J} illustrates the impact of point crossing on the current distribution $J$ with different junctions with MoM modeling.
Connecting the two areas with a single point shows a current distribution like the completely separated areas.
However, a negligible degree of overlap significantly changes current to be distributed like a loop antenna Fig.\ref{Fig_SQLoop_J}.
The radiation pattern is also influenced by the point crossing junctions (illustrated in Fig.\ref{Fig_SQLoop_FF}).
The current in Fig.\ref{Fig_SQLoop_J}.(a,b) is similar to the current distribution of a dipole with bent arms, so the radiation pattern is expected to be omnidirectional around $y$ axis.
However, for junction types of Fig.\ref{Fig_CornerShapes}(c-e), the current goes through the loop in the  $x-y$ plane.
This results in omnidirectional radiation around $z$ axis which is also validated by the simulated pattern in Fig.\ref{Fig_SQLoop_FF}(b).

\begin{figure}
\includegraphics[width=1\linewidth]{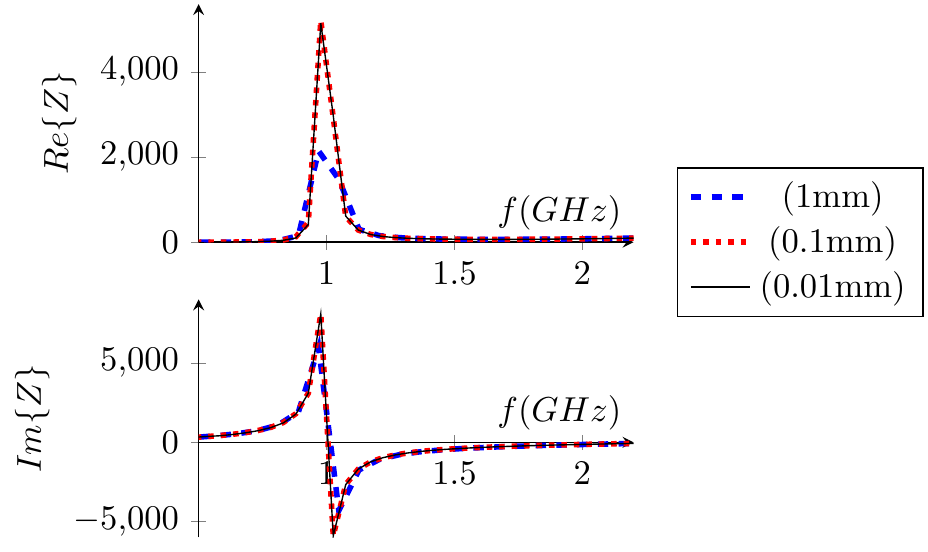}
\caption{Illustration of the impact of the width of the overlapping junction (bridge) over the impedance of the square loop antenna}
\label{Fig_JunctionWidth}
\end{figure}

The impact of the width of junction is shown in Fig.~\ref{Fig_JunctionWidth} where  the width of the bridge junction changes from $1$mm down to $0.01$mm. 
The impedance is almost the same in most of the frequency range, although it peaks to a higher value in the antiresonance region.
It is also seen from Fig.~\ref{Fig_SQLoop_J}(c-e) that the shape of the overlapping  connection does not make a significant difference to the results.


\section{Avoiding Point Crossing}
\label{Sec_AvoidPointCross}
It is desirable to avoid point crossing due to its impact on the performance of the antenna.
Avoiding such connections during optimization runs would need some schemes which ideally solves the point crossing issue with minimal modification of the geometry.
One plan would be to completely purge the conflicting cells \cite{Cismasu_2014_AWPL,Ohira_2004_TAP} from the structure.
This would avoid the issue, although the geometry modifications are not minimal.
In \cite{John_2007_AWPL}, conflicting cells were joined by small bridging junctions.
In this letter, we propose that one should consider both chamfering and bridging of the conflicting cells to properly explore the entire search space, and find optimum structures.

\begin{figure}
\subfigure[Antenna I]{
\includegraphics[width=0.4\linewidth]{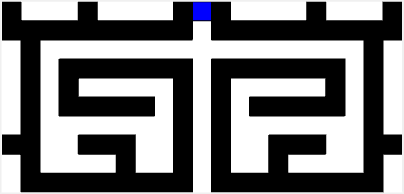}
}
\hfil
\subfigure[Antenna II]{
\includegraphics[width=0.4\linewidth]{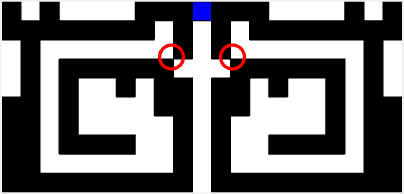}
\label{Fig_AntII}
}
\caption{Sample antennas without and with point crossing.
Note that the blue cell and red circles highlight the feed point, and the point crossing junctions, respectively.
}
\end{figure}

\textbf{Adaptive chamfering:}
One plan would be to identify the conflicting points in the structure. 
Then, conflicting cells are chamfered at the point contact cells alone (Fig.\ref{Fig_AntII}).

\textbf{Adaptive bridging:}
Another scenario should be to connect the neighbor cells.
As it is been illustrated in \cite{John_2007_AWPL}, adaptive bridging is also capable of providing optimum designs.

\begin{table}
\caption{Comparison of the simulation time by different EM tools} 
\label{Tab_SimulationTime}
\centering
\begin{tabular}{r|cccc}
\hline\hline   \rule{0pt}{2ex}
$t$(sec) & \textbf{FDTD} \cite{openEMS} & \textbf{FDTD} \cite{CST_MW}  & \textbf{MoM}\cite{FEKO} & \textbf{FEM}\cite{CST_MW}\\
\hline   \rule{0pt}{2ex} 
\textbf{Dipole}\\
Point Junction		& $\dagger$	&	 2		& 	33		&	72	 \\
Chamfer				& $\dagger$	&	 2		& 	35		& 	63	 \\
Bridge				& $\dagger$ &	 2		& 	38		& 	50	 \\
\hline   \rule{0pt}{2ex} 
\textbf{Square Loop}\\
Point Junction		&  $\dagger$&	 30		& 	94		&	47	 \\
Chamfer				&  $\dagger$&	 8		& 	62		& 	42	 \\
Bridge				&  $\dagger$&	 231	& 	62		& 	43	 \\
\hline   \rule{0pt}{2ex} 
\textbf{Antenna I}  &	37.9	&	15		&	42		&	67	\\
\hline  \rule{0pt}{2ex} \textbf{Antenna II}\\
Point Junction	  	&	349		&	30		&	58		&	293	\\
Chamfer 			&	59		&	14		&	57		&	63	\\
Bridge 				&	342		&	57		&	55		&	162 \\
\hline
\end{tabular}
\begin{flushleft}
\footnotesize{$\dagger$ The stucture was not simulated with this solver.}
\end{flushleft}
\end{table}

\begin{figure}
	\centering
	\includegraphics[height=0.45\linewidth]{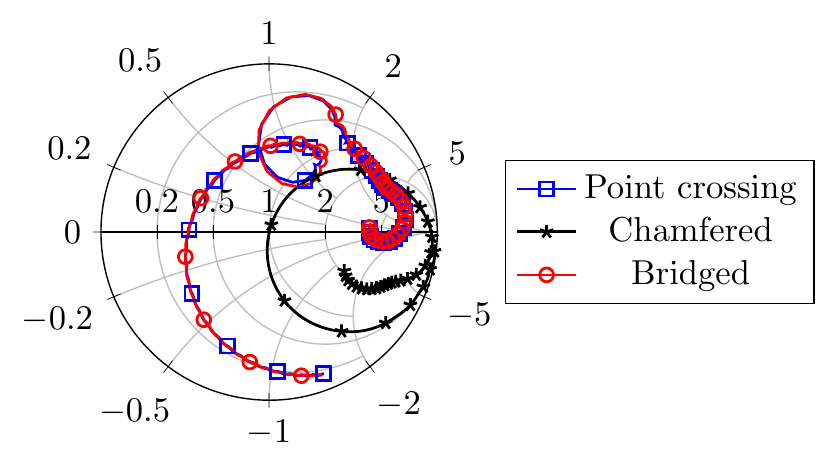}
	\caption{Impedance of the Antenna II on the Smith chart for different junctions using FDTD method.}
	\label{Fig_AntII_SmithChart}
\end{figure}

\section{Comparison of the simulation time}
\label{Sec_SimTime}

Because an optimizer runs hundreds and thousands of iterations to arrive to the optimized designs, the simulation time of a single run is one of the key factors in the optimization process.
It is, therefore, useful to see the impact of the proposed schemes on the simulation time.
Table.~\ref{Tab_SimulationTime} compares the simulation time of different solvers on dipole and loop antennas in Fig.~\ref{Fig_Dip_SQLoop}, \emph{Antenna I} which does not contain a point crossing, and \emph{Antenna II} which includes a point crossing and needs corner modification.
Each antenna was simulated with  different simulation packages: \emph{open-source FDTD } code\cite{openEMS}, \emph{commercial FDTD and FEM} code\cite{CST_MW}, and \emph{commercial MoM} code\cite{FEKO}.

As simulations are performed over a broad range of frequencies, the FDTD method is best suited to some antennas which have a good resonance ($Re \{Z\}$ not too small or too large e.g. dipole).
The simulation time of the dipole was not sensitive to the junction type.
However, chamfering point crossing junction would significantly impact on the simulation time of the square loop and antenna II.
As antennas with other junctions are not well-matched in the simulation frequency range (illustrated in Fig~.\ref{Fig_SQLoop_Imp} and Fig.~\ref{Fig_AntII_SmithChart}), FDTD simulations need more time for convergence.
Chamfered junctions show good matching near resonance, and their simulation times are significantly less.
Simulation time with MoM and FEM methods do not look to be sensitive to the junction type, as these methods are frequency domain methods.

\section{Additional remarks}
\begin{itemize}
\item We also considered the impact of point crossing on the FSS structures. 
	We simulated square loops as FSS, and observed a significant shift in the transmission and reflection bands of the structure (not included in this letter).
	We argue that in addition to antennas and FSS, point crossing should be avoided in other electromagnetic circuits (e.g. filters, couplers, etc.)

\item It should be noted that point crossing is not always problematic. 
	For instance, the measured optimized designs in \cite{Kern_2005_TAP,Xia_2013_PIER} showed a performance close to simulations.
	This might be due to insignificant flow of the current over the junction even if they are truly connected like Fig.~\ref{Fig_CornerShapes}(e).
	We argue that the effect of point crossing would be major when the junction prevents the flow of the current and so significantly changes the effective length.

\end{itemize}

\section{Conclusion}
In this letter, we illustrated that junctions with only one single point in common should be avoided in optimization algorithms.
We used MoM, FDTD, and FEM solvers to study and compare the impact of different junctions on the impedance, current distribution, and radiation pattern of the antennas.
We also compared the simulation time of different junctions.




\end{document}